\documentclass{article}

\usepackage[utf8]{inputenc}
\usepackage[T1]{fontenc}

\usepackage{mathptmx}       
\usepackage[scaled=0.92]{newpxtext}
\usepackage{courier}

\usepackage[a4paper,top=2cm,bottom=2cm,left=1.7cm,right=1.7cm]{geometry}

\usepackage{amsmath, amssymb, amsfonts, amsthm, mathtools}

\usepackage{graphicx}
\usepackage{subcaption}
\usepackage{caption}
\usepackage[table,xcdraw]{xcolor}
\usepackage{array}
\usepackage{colortbl}

\usepackage{braket}
\usepackage{quantikz}

\usepackage{multicol}
\usepackage{url}
\usepackage{hyperref}
\usepackage{tikz}
\usetikzlibrary{arrows.meta,positioning,shapes.geometric,fit,calc}

\usepackage{mathrsfs}
\usepackage{appendix}
\usepackage{parskip}

\usepackage{pgfplots}
\pgfplotsset{compat=1.17}
\usepackage{authblk}

\newtheoremstyle{thmstyleone}
  {3pt}{3pt}{\itshape}{}{\bfseries}{.}{0.5em}{}
\newtheoremstyle{thmstyletwo}
  {3pt}{3pt}{}{}{\bfseries}{.}{0.5em}{}
\newtheoremstyle{thmstylethree}
  {3pt}{3pt}{}{}{\bfseries}{.}{0.5em}{}

\theoremstyle{thmstyleone}

\theoremstyle{thmstyletwo}

\theoremstyle{thmstylethree}

\title{\textbf{From Period Finding to Lattice Sampling: Experimental Insights into Shor’s and Regev’s Factoring Algorithms}}

\author[1]{Daniela Falcó Pomares}
\author[2]{Arturo Rodríguez-Almazán}
\author[1]{Guillermo Rivas}
\author[2,4]{Ricardo S. Alonso}

\affil[1]{AIR Institute, Av. Santiago Madrigal, Salamanca, 37003, Castilla y León, Spain}
\affil[2]{AIR Institute, Carr. de la Aldehuela, Zamora, 49022, Castilla y León, Spain}
\affil[3]{AIR Institute, Av. Santiago Madrigal, Salamanca, 37003, Castilla y León, Spain}
\affil[4]{International University of La Rioja, Av. de la Paz, Logroño, 26006, La Rioja, Spain}

\date{}

\begin{document}

\maketitle
\begin{abstract}
Quantum algorithms for integer factorization represent one of the most prominent applications of quantum computation, with far reaching implications for modern cryptography. While Shor’s algorithm provides a polynomial time solution in the ideal quantum model, its practical implementation is severely constrained by the limitations of current noisy intermediate scale quantum (NISQ) hardware. These constraints have motivated the exploration of alternative factoring algorithms with different structural and resource trade-offs.

In this work, we present an experimental study of Regev’s quantum factoring algorithm, implemented on real quantum hardware, and compare its behavior with that of Shor’s algorithm under analogous conditions. Focusing on the case $N = 15$, we execute both algorithms on the QMIO quantum computer at the Centro de Supercomputación de Galicia (CESGA) and contrast the results with one of IBM’s open access quantum computers and ideal simulations. This parallel execution enables a low level comparison of the two algorithms, highlighting how their respective quantum implementations interact with hardware noise, limited circuit depth, and finite sampling.

Our analysis emphasizes the different ways in which Shor’s and Regev’s algorithms encode arithmetic structure into quantum states through Fourier sampling in one and higher dimensions, respectively, and how these differences manifest in experimental outcomes. Although neither algorithm demonstrates a practical advantage in the small $N$ regime, the results provide insight into their relative robustness and failure modes on contemporary quantum devices. This study illustrates the value of experimental benchmarking of alternative quantum factoring algorithms as a means of understanding the practical implications of algorithmic design choices in the NISQ era.
\end{abstract}

\vspace{0.5 cm}


\section{Introduction and Background}\label{sec:Introduction}

Integer factorization has long occupied a central position in number theory and, more recently, in public-key cryptography. In particular, the security of widely deployed cryptographic schemes such as RSA relies on the presumed computational intractability of factoring large composite integers using classical algorithms. This assumption underpins a substantial fraction of modern digital security infrastructures.

The advent of quantum computing has fundamentally challenged this premise. Shor’s seminal algorithm demonstrated that integer factorization and discrete logarithmic computation can be performed in polynomial time on a quantum computer, implying that RSA and elliptic-curve cryptography would be rendered insecure in the presence of sufficiently large and reliable quantum hardware \cite{Shor_1997, ekera_2016}. As a consequence, classical public-key cryptography is widely considered to face an existential threat from quantum algorithms \cite{mosca_2018, raheman2022quantumcyber}.

In recent years, advances in quantum resource estimation have further sharpened this concern. Notably, improved fault tolerant constructions and error correction analyses suggest that factoring a $2048$-bit RSA modulus could be achieved using on the order of one million noisy physical qubits sustained for approximately one week, representing a dramatic reduction compared to earlier estimates that required tens of millions of qubits \cite{gidney2021, spaceEfficientFactoring}. These developments have significantly accelerated the anticipated timeline for the practical relevance of quantum attacks on cryptographic systems.

Since its original proposal, Shor’s algorithm has undergone extensive theoretical and practical refinement. Numerous studies have focused on reducing circuit depth and qubit requirements through optimized arithmetic constructions, alternative modular exponentiation schemes, and improved phase estimation techniques. Notable examples include low-qubit implementations of Shor’s algorithm, refinements based on mixed state models, and the development of efficient quantum addition and multiplication circuits, which form the backbone of modular arithmetic in quantum algorithms \cite{beauregard2002, parker2000}. In parallel, numerous experimental demonstrations and simulations have explored the feasibility of Shor’s algorithm at small problem sizes, typically for $N=15$ or $N=21$, providing valuable insights into the effects of noise, decoherence, and hardware constraints.

Despite the profound theoretical implications of quantum algorithms for factorization, their practical realization remains constrained by the limitations of current quantum hardware. Present day quantum processors operate in the so called noisy intermediate scale quantum (NISQ) regime, characterized by a limited number of qubits, restricted qubit connectivity, and non-negligible gate and measurement errors \cite{Preskill_2018}. In this setting, the execution of deep quantum circuits poses a significant challenge, as errors accumulate rapidly and can severely degrade the reliability of the computed results.
These practical considerations have motivated the exploration of alternative quantum algorithms whose structures may be better suited to near term or intermediate scale quantum devices. In particular, algorithms that trade circuit depth for repeated executions combined with classical post-processing have attracted increasing attention. In this context, Regev introduced a quantum factoring algorithm that can be interpreted as a higher dimensional generalization of the period finding paradigm underlying Shor’s algorithm \cite{regev2023, regevOverview}. While Shor’s algorithm can be viewed as recovering a one-dimensional periodic lattice, Regev’s approach extends this idea to higher dimensional lattices, reformulating the factorization problem as a lattice sampling task followed by classical recovery of short lattice vectors.

Following its initial proposal, Regev’s algorithm has been the subject of several theoretical and methodological investigations. These works have addressed its unconditional correctness, probabilistic guarantees, extensions to related problems such as discrete logarithms, and the interplay between quantum sampling and classical lattice reduction techniques \cite{pilatte2023, implementationRegev}. More recently, implementation oriented studies have explored the practical realization of Regev’s algorithm and compared its performance with Shor’s algorithm, with particular emphasis on runtime behavior, parameter sensitivity, and classical reconstruction efficiency \cite{pawlitko2025implementation}. From a resource theoretic perspective, Regev’s algorithm offers an alternative trade off: it employs quantum circuits of asymptotically smaller depth at the cost of requiring multiple independent executions and a more involved classical reconstruction stage.

A central hypothesis of this paper is that Shor’s and Regev’s algorithms exhibit qualitatively different robustness profiles on NISQ devices because they encode arithmetic information in different ways within the quantum state. Shor’s algorithm aims to concentrate useful information about the hidden period into a small number of sharp phase estimation peaks. In contrast, Regev’s algorithm distributes information across a higher dimensional set of correlated Fourier sampling outcomes whose geometric bias becomes exploitable only after aggregating multiple samples in the classical post-processing stage. This perspective will be used to interpret how hardware noise manifests in the observed measurement histograms and why the two algorithms fail in different ways even when their compiled depths are comparable.

However, existing experimental analyses primarily focus on performance metrics or single platform evaluations. To the best of our knowledge, a systematic multi backend comparison of Shor’s and Regev’s algorithms under tightly controlled and comparable compilation and execution conditions adapted to each hardware architecture has not yet been reported. In particular, the way in which both algorithms encode arithmetic structure into quantum measurement distributions, and how these structural differences interact with realistic hardware noise across independent quantum devices, remains insufficiently characterized at the experimental level.

The objective of this work is to present a controlled experimental study of Regev’s factoring algorithm applied to the case \(N = 15\), executed on real quantum hardware and compared with idealized classical simulations. Specifically, we implement the algorithm on the QMIO quantum computer at the \textsc{CESGA} and contrast the obtained results with those produced by one of IBM’s open-access quantum processors, as well as with noiseless simulations. In order to enable a direct and low-level comparison, Shor’s algorithm is executed under analogous compilation and execution conditions on the same platforms. By analyzing the behavior of both algorithms in the presence of realistic hardware noise and resource constraints, this study aims to shed light on their relative robustness, distinct failure modes, and structural differences in the NISQ regime.

The contributions of this work are:
\begin{itemize}
  \item \textbf{Controlled experimental benchmark (same instance, comparable settings).}
  We implement and execute Shor’s and Regev’s factoring algorithms for the benchmark instance \(N=15\) under comparable compilation and execution conditions, enabling a direct comparison of their measurement statistics and failure modes on NISQ hardware.

  \item \textbf{Multi backend evaluation.}
  We run the implementations on the QMIO quantum computer at \textsc{CESGA} and on an IBM open access quantum processor, and we compare both against noiseless simulations as a reference baseline.

  \item \textbf{Information encoding interpretation.}
  We interpret the experimental outcomes through an information concentration vs.\ information distribution perspective, explaining qualitatively different noise induced degradations in the observed measurement distributions for Shor and Regev.

  \item \textbf{Reproducibility oriented reporting.}
  We report the algorithmic parameter choices for the \(N=15\) instances together with the compilation/execution pipeline used to obtain the experimental distributions.
\end{itemize}

The remainder of the paper is structured as follows.
In Section~2 we review the quantum factoring algorithms considered in this work,
focusing on the structural differences between Shor’s period finding approach
and Regev’s lattice based formulation.

Section~3 presents the analytical and experimental study for the benchmark
instance $N=15$. We first describe the experimental methodology and the
quantitative metrics used to compare measurement distributions, and then
report the results obtained for Shor’s and Regev’s algorithms on simulation
and real quantum hardware, followed by a comparative discussion of their
behavior and a summary of the main limitations of the present study.
\section{Quantum factoring algorithms: Shor vs Regev}\label{ShorvsRegev}

Both Shor’s and Regev’s algorithms reduce integer factorization to the recovery of hidden
algebraic structure using a combination of a quantum sampling routine and a classical
post-processing stage. Although the two algorithms are based on closely related ideas, they encode
arithmetic information in different but structurally related ways, leading to distinct resource
requirements and different behavior on noisy quantum hardware.

In both cases, the computation can be decomposed into two main parts:

\begin{itemize}
\item A quantum subroutine that produces measurement outcomes containing partial
information about the hidden structure, and
\item a classical reconstruction stage that extracts non-trivial factors of $N$
from these outcomes.
\end{itemize}

The main difference lies in the dimensionality of the hidden structure that must be
recovered. Shor’s algorithm reduces factoring to the recovery of a one-dimensional
period, which can be interpreted as the simplest case of a lattice of relations,
whereas Regev’s algorithm generalizes this idea to higher-dimensional lattices of
multiplicative relations that must be reconstructed from several independent samples.

\subsection{General structure of quantum factoring algorithms}

Let $N$ be an odd composite integer. Both Shor’s and Regev’s algorithms follow the same general paradigm. In each case,
the factorization problem is reduced to the recovery of hidden algebraic structure,
which is accessed through a quantum sampling procedure and reconstructed using
classical post-processing.

In Shor’s algorithm the relevant structure is a one-dimensional period, which can be
viewed as the simplest instance of a lattice of relations, whereas in Regev’s algorithm
the hidden structure is a higher dimensional lattice of multiplicative relations.
As a consequence, the quantum subroutine and the classical reconstruction stage play analogous roles
in both algorithms, but operate on structures of different dimensionality rather than on fundamentally
different mathematical objects.

\begin{center}
\begin{tabular}{c|c|c|c}
Algorithm & Hidden structure & Quantum step & Classical step \\
\hline
Shor & Period $r$ / $1$D Lattice & Fourier Sampling & Continued Fractions \\
Regev & Lattice $L$ & Fourier Sampling & Lattice Reduction
\end{tabular}
\end{center}

This distinction will be important for the experimental comparison, since the way
information is encoded in the quantum state determines how noise affects the
observed measurement distributions.

\subsection{Shor’s algorithm}

Shor’s algorithm reduces integer factorization to order finding in the multiplicative
group $\mathbb{Z}_N^*$. For a randomly chosen $a \in \mathbb{Z}_N^*$, the goal is to
determine the smallest integer $r > 0$ such that

\[
a^r \equiv 1 \pmod N.
\]

If $r$ is even and $a^{r/2} \not\equiv -1 \pmod N$, non-trivial factors of $N$
are obtained from

\[
\gcd(a^{r/2} \pm 1, N).
\]

The quantum part of the algorithm applies quantum phase estimation to the modular
multiplication operator

\[
\mathrm{U}_{a,N} |w\rangle = |a \cdot w \bmod N\rangle,
\]

which can be interpreted as a one-dimensional Fourier sampling procedure over the
hidden period. The measurement outcomes approximate rational values of the form $s/r$,
from which the order $r$ is recovered using continued fractions as part of the classical
post-processing, and the factors of $N$ are then obtained.

In this formulation, the relevant information about the hidden period, which can be
viewed as a one-dimensional lattice of relations, is concentrated into a small number
of peaks in the Fourier domain. When these peaks are well resolved, the reconstruction
is straightforward. However, this concentration also makes the method sensitive to phase
errors and decoherence, since the useful information is localized in a small subset of
measurement outcomes.

\subsection{Regev’s algorithm}

Regev’s algorithm generalizes one-dimensional period finding to the recovery of a
higher dimensional periodic structure described by a lattice of multiplicative
relations. Instead of determining a single period, the algorithm identifies
integer relations between modular exponentiations, which form a lattice whose
short vectors encode non-trivial factors of $N$.

Let $a_1, \dots, a_d \in \mathbb{Z}_N^*$ and consider integer vectors
$z = (z_1,\dots,z_d) \in \mathbb{Z}^d$ satisfying

\[
\prod_{i=1}^d a_i^{z_i} \equiv 1 \pmod N.
\]

These relations form a lattice

\[
L =
\left\{
z \in \mathbb{Z}^d :
\prod_{i=1}^d a_i^{z_i} \equiv 1 \pmod N
\right\}.
\]

Given a non-trivial vector $z \in L$, define

\[
b = \prod_{i=1}^d a_i^{z_i} \bmod N.
\]

Such a vector produces a square root of unity modulo $N$, and a non-trivial factor
of $N$ can then be obtained from

\[
\gcd(b \pm 1, N).
\]

The quantum subroutine prepares a Gaussian superposition over $\mathbb{Z}^d$,
applies modular exponentiation in superposition, and performs a Fourier transform.
The resulting measurement outcomes provide noisy approximations of vectors in the
dual lattice $L^*$.

The classical post-processing stage reconstructs short vectors in $L$ from these
samples using lattice-reduction techniques such as LLL. In contrast to Shor’s
algorithm, the relevant information is not concentrated in a single measurement
outcome but distributed across multiple samples, which must be combined to recover
the lattice relations.

\subsection{Structural comparison}\label{subsec:structural}

Although both algorithms ultimately reduce factorization to the recovery of hidden
algebraic structure, they differ mainly in the dimensionality and representation of
this structure in the quantum state.

Shor’s algorithm reduces factoring to the determination of a one-dimensional period,
which can be viewed as the simplest instance of a lattice of relations. The quantum
phase estimation procedure concentrates the relevant information into a small number
of peaks in the Fourier domain, from which the period can be recovered efficiently
using continued fractions. This concentration makes the reconstruction simple in the
ideal setting, but also makes the algorithm sensitive to noise and loss of coherence,
since the useful information is localized in a small subset of measurement outcomes.

Regev’s algorithm generalizes this idea to the recovery of a higher dimensional periodic
structure described by a lattice of multiplicative relations. The quantum subroutine
produces samples close to the dual lattice, and the final reconstruction amounts to
finding short vectors in the corresponding lattice using classical lattice reduction
techniques. In this case, the relevant information is not concentrated in a single
spectral peak but distributed across multiple correlated measurement outcomes whose
geometric structure becomes exploitable only after aggregation.

These different representations lead to distinct resource trade-offs. Shor’s algorithm
requires deeper quantum circuits dominated by modular exponentiation, but relatively
simple classical post-processing, whereas Regev’s algorithm uses shallower quantum
circuits at the cost of repeated sampling and a more involved classical reconstruction.

This distinction is particularly relevant in the NISQ regime, where circuit depth,
noise, and sampling limitations strongly influence the observed behavior. The
experimental comparison presented in this work is motivated by this difference between
information concentration and information distribution in the two algorithms.
\section{Analytical and Numerical Study for the Case \texorpdfstring{$N=15$}{N=15}}\label{sec:CaseStudy}

\subsection{Experimental methodology}\label{subsec:methodology}

All circuits were implemented using Qiskit and are composed of three main building blocks:
state preparation, modular exponentiation in superposition, and a Fourier transform
stage followed by measurement. While both Shor’s and Regev’s algorithms
share this high level structure, they differ in the dimensionality and representation of the
arithmetic relations encoded in the quantum state.

\paragraph{Arithmetic subroutines and circuit construction.}

In general, modular arithmetic in quantum factoring algorithms is implemented using
reversible ripple-carry adders, modular addition, and controlled modular multiplication
subroutines, which are combined to construct modular exponentiation circuits.
Such generic constructions allow a faithful implementation of both Shor’s and Regev’s
algorithms, but typically lead to circuits with substantial depth even for small values
of $N$.

For the specific benchmark $N=15$, instead of using fully generic arithmetic circuits,
we employed specialized \emph{ad hoc} modular exponentiation constructions.
These circuits implement the required modular action as shallow permutation-based
transformations, significantly reducing circuit depth.

This design choice allows us to focus the experimental analysis on the behavior of the
Fourier sampling stage under hardware noise, rather than on the performance of large
arithmetic subcircuits, which would dominate the depth in a fully general implementation.

\paragraph{Algorithm parameters.}

For Shor’s algorithm, we consider the instance $N=15$ with base $a=2$, which has
order $r=4$ modulo $15$. The control register size was fixed to $k=3$ qubits,
sufficient to resolve the relevant phase information for this instance.
Post-processing is performed using the standard continued fraction method to
recover the order and extract non-trivial factors via
$\gcd(a^{r/2}\pm 1,N)$.

For Regev’s algorithm, we consider a two-dimensional instance ($d=2$) with
parameters
\(
a_1 = 4\), \(a_2 = 49\). These values define the modular relations used to construct the lattice of
multiplicative dependencies. The Gaussian width parameter was set to
\( 
R = \exp(dC)\) with \( C = 0.5,\) which determines the discretizations of the sampling register through
\(
k=\left\lfloor \log_2(\sqrt{d}\,R)\right\rfloor + 1
\)
and the corresponding noise parameter
\(
\delta=\sqrt{d}/\sqrt{2}\,R,
\) used in the classical post-processing stage.
Measurement outcomes are parsed into $d$ integer coordinates and ranked by
frequency before being passed to the lattice reduction routine.

\paragraph{Backends, compilation and shot budgets.}

Circuits were compiled using the Qiskit transpiler, with compilation parameters
adjusted for each backend in order to obtain the best trade-off between circuit
depth, gate count, and hardware constraints. Due to differences in device geometry
and native gate sets, compilation settings were adapted separately for IBM Quantum
and QMIO processors.

All hardware executions were performed with $4096$ shots. In addition, ideal
(noiseless) simulations were used as a reference to characterize the expected
measurement distributions in the absence of hardware noise.

\subsection{Quantitative metrics and success criteria}\label{subsec:metrics}

To complement the qualitative inspection of measurement counts and their
histogram representations, we introduce quantitative metrics that characterize
the structure of the empirical shot distributions obtained from each execution.
Since the experimental data consist of aggregated shot counts, we focus on
distribution based metrics that can be computed directly from the observed
probability distributions without requiring full classical reconstruction
for each run.

These metrics allow a reproducible comparison across algorithms, backends,
and compilation settings, and provide a quantitative way to analyze the
concentration versus distribution behavior discussed in Section~\ref{subsec:structural}.
Whenever relevant, we distinguish between metrics that quantify the spread
of the observed distributions and metrics that relate to the success of the
classical reconstruction stage.

\paragraph{Distribution based comparison.}
Both Shor’s and Regev’s algorithms encode arithmetic information into
measurement distributions whose structure is essential for the success of
the classical reconstruction stage. In Shor’s algorithm, useful information
is typically concentrated in a small number of spectral peaks, whereas in
Regev’s algorithm information is distributed across several correlated
outcomes in a higher-dimensional sampling space. Hardware noise tends to
flatten these structures, making it necessary to quantify how strongly the
probability mass remains concentrated.

Because the available data consist of shot counts collected for each
backend, we evaluate the performance of the algorithms through metrics
that depend only on the empirical probability distribution.

\paragraph{Shannon entropy.}
Given an empirical distribution $p$ obtained from the measurement shots,
we compute the Shannon entropy
\[
H(p) = -\sum_x p(x)\log p(x),
\]
which measures the overall dispersion of the distribution.
Low entropy indicates that the probability mass is concentrated on a small
set of outcomes, while high entropy corresponds to a flatter distribution.
This metric is useful for characterizing how strongly the measurement
distribution deviates from the ideal structure expected for each algorithm,
and therefore provides a quantitative indicator of the effect of hardware
noise on the encoded arithmetic information.

\paragraph{Effective support size.}
We also compute the effective support size
\[
N_{\mathrm{eff}}(p) = \frac{1}{\sum_x p(x)^2},
\]
also known as the inverse participation ratio.
This quantity estimates the number of outcomes that contribute
significantly to the distribution.
For Shor’s algorithm, an ideal execution typically exhibits a small
effective support corresponding to a few dominant peaks, while for
Regev’s algorithm the distribution is generally more spread even in the
noiseless case for the instances considered here.
An increase of $N_{\mathrm{eff}}$ under hardware noise indicates that the
distribution becomes more uniform and that the encoded structure is being
degraded.

\paragraph{Top-$K$ mass.}
To directly quantify the concentration of probability on the most relevant
outcomes, we report the top-$K$ mass
\[
M_K(p)=\sum_{x\in \text{Top-}K} p(x),
\]
with $K=4$ in all experiments.
This metric measures the fraction of total probability contained in the
four most frequent measurement results.
For Shor’s algorithm, the ideal distribution for the instance considered
here is expected to be strongly concentrated on four dominant peaks, so
$M_4$ should be close to one in the noiseless case.
For Regev’s algorithm, the ideal distribution is typically more spread,
but for the parameters used in this work a small number of outcomes may
still carry a significant fraction of the probability, making $M_4$ a
useful indicator of how much the sampled lattice structure is degraded
by hardware noise.

Together, the quantities $H(p)$, $N_{\mathrm{eff}}(p)$, and $M_4(p)$ provide
a consistent way to quantify how the information encoded in the quantum
state changes across different hardware platforms, and they serve as
numerical indicators of the concentration versus distribution viewpoint
introduced in Section~\ref{subsec:structural}.

The numerical values obtained for these quantities in our experiments
are reported in the Table~\ref{tab:metrics} and will be used to support
the qualitative discussion of the measurement distributions and to
provide a consistent comparison between Shor’s and Regev’s algorithms
across different hardware platforms.



\begin{table}[t]
\centering
\caption{Quantitative distribution metrics for Shor’s and Regev’s algorithms for $N=15$.}
\label{tab:metrics}
\begin{tabular}{llcccc}
\hline
Backend & Algorithm & $H(p)$ &$N_{eff}$ &$M_4(p)$ \\
\hline
Simulation & Shor & 1.386 & 3.998 & 1.0 \\
IBM Quantum & Shor & 1.662 & 4.678 & 0.921 \\
QMIO & Shor & 2.068 & 7.827 & 0.564 \\
\hline
Simulation & Regev & 1.074 & 2.343 & 0.944 \\
IBM Quantum & Regev & 2.205 & 6.506 & 0.645 \\
QMIO & Regev & 2.766 & 15.795 & 0.29 \\
\hline
\end{tabular}
\end{table}

\subsection{Results for Shor’s algorithm}\label{subsec:shor_results}

We begin by analyzing the experimental behavior of Shor’s algorithm for the instance $N=15$ with base $a=2$. As discussed in Section~\ref{ShorvsRegev}, the quantum subroutine aims to extract phase information corresponding to the order $r=4$ through a phase-estimation (Fourier sampling) procedure, followed by classical post-processing based on continued fractions.

\paragraph{Idealized behavior.}
In the noiseless simulation setting, the measurement outcomes of the control register exhibit a highly structured distribution. The probability mass is concentrated on a small subset of outcomes corresponding to phases that approximate rational values of the form $s/r$, with $r=4$. This behavior is consistent with the theoretical expectation for the phase-estimation subroutine and provides a reference baseline against which hardware results can be compared.

This strong concentration is also reflected in the quantitative metrics reported in Table~\ref{tab:metrics}. The entropy is low ($H=1.386$), the effective support size is close to four ($N_{\mathrm{eff}}\approx4$), and the top-$4$ mass is equal to one, indicating that the probability is entirely concentrated on the expected phase estimation peaks.

\paragraph{IBM Quantum results.}
When executed on IBM Quantum hardware, the measurement distribution deviates from the ideal case due to gate imperfections, decoherence, and measurement noise. Nevertheless, the distribution retains partial structure: a small number of outcomes appear with significantly higher frequency than the rest. These dominant outcomes remain compatible with rational approximations to multiples of $1/r$, so that the classical continued fraction reconstruction would still be expected to recover the correct order in a non-negligible fraction of cases.

This behavior indicates that, for the circuit depths considered here, the phase estimation signal is degraded but not completely destroyed on IBM hardware. The observed broadening of peaks reflects the sensitivity of Shor’s algorithm to accumulated phase errors, while the persistence of dominant outcomes suggests a degree of robustness at this small scale. Consistently with this observation, Table~\ref{tab:metrics} shows a moderate increase in entropy ($H=1.662$) and effective support size ($N_{\mathrm{eff}}\approx4.68$), while the top-$4$ mass remains high ($M_4=0.921$), confirming that the distribution is still largely concentrated on a few dominant results.

\paragraph{QMIO results.}
In contrast, executions on the QMIO device yield measurement distributions that are substantially flatter. The probability mass is spread more uniformly across outcomes, and no clearly dominant peaks corresponding to the expected phase information are observed. As a result, the success of the classical post-processing stage becomes highly sensitive to threshold choices and statistical fluctuations.

This flattening of the distribution illustrates a characteristic failure mode of Shor’s algorithm on noisier hardware: the loss of sharp spectral features in the control register. Since the algorithm relies on the emergence of well-defined one-dimensional frequency peaks, even moderate levels of noise can obscure the underlying periodic structure. The quantitative metrics confirm this behavior: the entropy increases to $H=2.068$, the effective support size grows to $N_{\mathrm{eff}}\approx7.83$, and the top-$4$ mass drops to $M_4=0.564$, indicating that the probability is no longer concentrated on a small set of phase-estimation outcomes.

\begin{figure}[h]
\centering
\includegraphics[width=\textwidth]{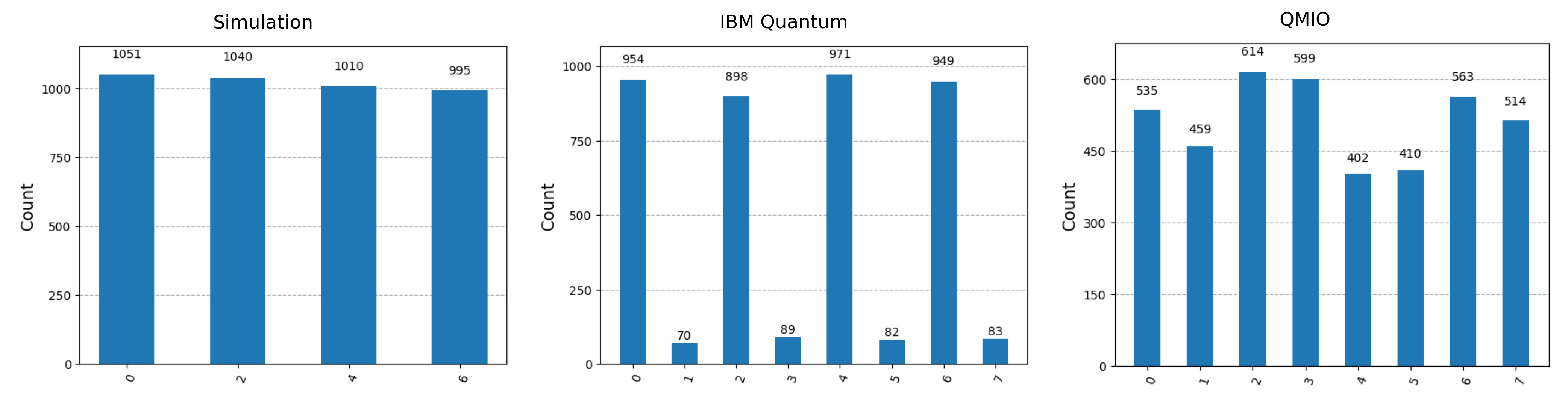}
\caption{
Measurement distributions for Shor’s algorithm ($N=15$, $a=2$, $k=3$).
From left to right: ideal simulation, IBM Quantum hardware results, and QMIO hardware results.
The ideal distribution exhibits well-defined peaks corresponding to the order $r=4$.
On real hardware, noise and decoherence progressively broaden and flatten the distribution, with a more pronounced loss of spectral structure observed on the QMIO device.
}
\label{fig:shor_results}
\end{figure}

\paragraph{Summary.}
Overall, these results highlight the dependence of Shor’s algorithm on the integrity of the phase estimation subroutine. Both the distributions shown in Fig.~\ref{fig:shor_results} and the numerical metrics in Table~\ref{tab:metrics} show that hardware noise progressively broadens the spectral peaks, leading to a loss of concentration in the measurement distribution and a reduced reliability of the classical reconstruction stage.

\subsection{Results for Regev’s algorithm}\label{subsec:regev_results}

We now analyze the experimental behavior of Regev’s factoring algorithm for the instance $N=15$ with dimension $d=2$ and parameters specified in Section~\ref{subsec:methodology}. In contrast to the one-dimensional case considered in Shor’s algorithm, Regev’s algorithm produces samples that approximate elements of a dual lattice through multidimensional Fourier sampling. As a result, the expected measurement distributions are inherently more dispersed.

\paragraph{Idealized behavior.}
In the noiseless simulation setting, the measurement outcomes exhibit a structured but more distributed pattern. Although the distribution is not sharply peaked at a single outcome, a small set of vectors appears with significantly higher frequency. These dominant vectors are consistent with noisy approximations of elements of the dual lattice and provide suitable input for the classical lattice based post-processing stage.

This behavior reflects the intended design of Regev’s algorithm: rather than concentrating probability mass into narrow peaks, the algorithm distributes information across multiple correlated outcomes, relying on statistical bias and geometric structure. This difference with respect to Shor’s algorithm is also visible in the quantitative metrics reported in Table~\ref{tab:metrics}. Even in the noiseless simulation, the entropy remains relatively low ($H=1.074$) and the effective support size is slightly above two ($N_{\mathrm{eff}}\approx2.34$), indicating that the probability is still concentrated on a limited subset of outcomes. Nevertheless, the top-$4$ mass remains high ($M_4=0.944$), showing that a small subset of dominant vectors still carries most of the information.

\paragraph{IBM Quantum results.}
When executed on IBM Quantum hardware, the measurement distributions become more dispersed compared to the theoretical reference. Nevertheless, a subset of outcomes still appears with noticeably higher frequency than the background. While individual samples are noisier, the presence of these dominant vectors indicates that partial geometric information about the underlying lattice survives the effects of hardware noise. In particular, some measurement outcomes may be partially consistent with the expected lattice directions even when others are significantly perturbed.

In this regime, the classical post-processing stage remains compatible with the most frequent measurement outcomes, although its success becomes more sensitive to the choice of parameters and the number of available samples. Compared to Shor’s algorithm, the degradation manifests as an increased spread around the expected geometric structure. Consistently with this observation, Table~\ref{tab:metrics} shows a clear increase in entropy ($H=2.205$) and effective support size ($N_{\mathrm{eff}}\approx6.51$), together with a decrease of the top-$4$ mass ($M_4=0.645$), indicating that the probability distribution becomes significantly more spread while still retaining a subset of dominant outcomes.

\paragraph{QMIO results.}
On the QMIO device, the measurement outcomes are significantly more uniformly distributed across a large number of vectors. No small subset of dominant outcomes clearly stands out from the background. As a consequence, the geometric signal produced by the quantum subroutine is strongly diluted by noise, and the classical lattice-reduction stage receives less informative input.

This behavior illustrates a characteristic failure mode of Regev’s algorithm on noisier hardware. Noise disperses the higher-dimensional Fourier samples, effectively obscuring the geometric correlations required for reliable lattice reconstruction. The quantitative metrics confirm this strong dispersion: the entropy increases to $H=2.766$, the effective support size grows to $N_{\mathrm{eff}}\approx15.80$, and the top-$4$ mass drops to $M_4=0.29$, showing that the probability is spread over many outcomes and that the dominant geometric structure is largely lost.

\begin{figure}[h]
\centering
\includegraphics[width=\textwidth]{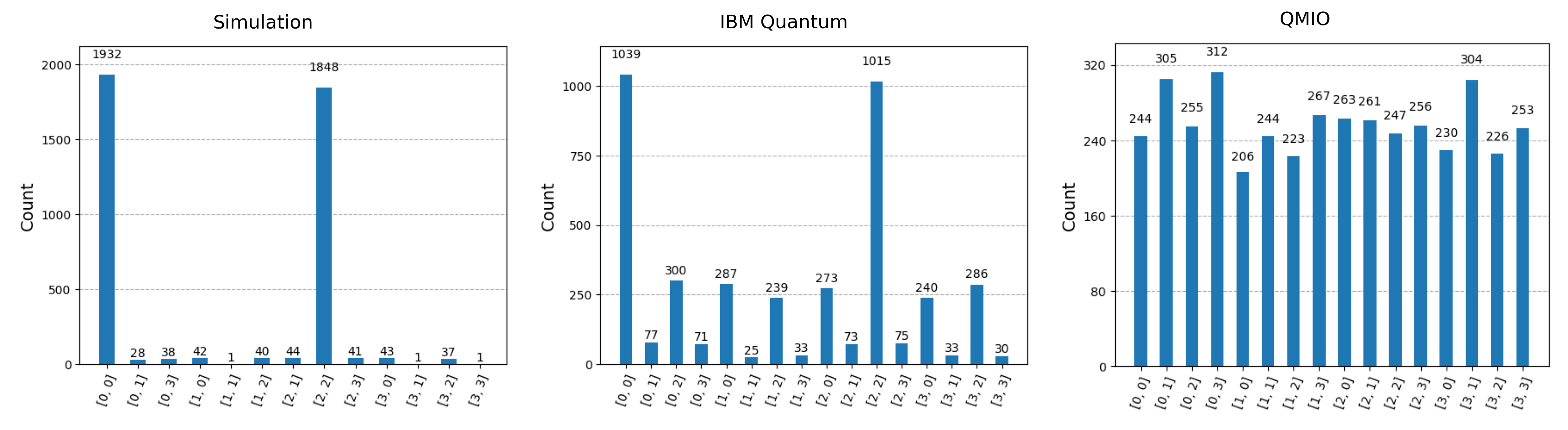}
\caption{
Measurement distributions for Regev’s factoring algorithm ($N=15$, $d=2$).
From left to right: ideal simulation, IBM Quantum hardware results, and QMIO hardware results.
In the ideal case, the distribution exhibits a small set of dominant outcomes consistent with approximate dual lattice sampling.
On real hardware, noise and decoherence lead to a more dispersed distribution, with a stronger dilution of geometric structure observed on the QMIO device.
}
\label{fig:regev_results}
\end{figure}

\paragraph{Summary.}
These effects are visible both in the distributions shown in Fig.~\ref{fig:regev_results} and in the quantitative metrics of Table~\ref{tab:metrics}. Compared to Shor’s algorithm, Regev’s method exhibits more distributed measurement patterns, and hardware noise further increases this dispersion, reducing the amount of useful geometric information available to the classical reconstruction stage.

\subsection{Comparative discussion: Shor vs Regev}\label{subsec:comparison}

Beyond qualitative differences in measurement distributions, Shor’s and Regev’s algorithms also exhibit distinct depth profiles after compilation, as summarized in Table~\ref{tab:resources}. In addition, the quantitative distribution metrics reported in Table~\ref{tab:metrics} provide a numerical characterization of how the information encoded in the quantum state is affected by hardware noise.

The experimental results presented above allow for a direct comparison between Shor’s and Regev’s factoring algorithms under similar hardware constraints. Although both algorithms target the same computational problem, they encode arithmetic structure into quantum states in different but closely related ways, leading to distinct experimental behaviors and failure modes.

Shor’s algorithm concentrates information about the hidden period into a one-dimensional spectral signal. This behavior is reflected in the noiseless metrics of Table~\ref{tab:metrics}, where the entropy is low and the effective support size is close to four, indicating that the probability mass is concentrated on a small number of phase estimation peaks. When this structure is preserved, as partially observed on IBM Quantum hardware, the classical post-processing stage remains compatible with recovering the correct order even in the presence of moderate noise. However, this concentration of information also makes the algorithm sensitive to phase errors and decoherence. As observed on the QMIO device, the entropy increases and the effective support grows significantly, while the top-$4$ mass decreases, showing that the spectral peaks are broadened and the underlying periodic structure becomes harder to resolve.

In comparison, Regev’s algorithm distributes information across a higher dimensional measurement space. Rather than relying on the emergence of narrow peaks, it exploits statistical bias and geometric correlations among multiple approximate samples from the dual lattice. This behavior is visible in the noiseless metrics of Table~\ref{tab:metrics}, where the entropy remains relatively low but the effective support reflects a more distributed probability pattern compared to Shor’s ideal case. Under hardware noise, these quantities increase further and the top-$4$ mass decreases, reflecting the progressive dilution of the geometric structure required for lattice reconstruction.

From an implementation perspective, Regev’s algorithm offers certain advantages in terms of circuit depth, as evidenced by the shallower circuits obtained after transpilation on both IBM Quantum and QMIO platforms (Table~\ref{tab:resources}). Nevertheless, this reduction in depth does not translate directly into increased robustness on current hardware. In particular, the multidimensional Fourier sampling stage and the preparation of discrete Gaussian states introduce additional sources of noise sensitivity, which can outweigh the benefits of reduced arithmetic complexity at small problem sizes. The metrics in Table~\ref{tab:metrics} confirm that, despite the smaller circuit depth, the measurement distributions for Regev become more rapidly dispersed under noise than those of Shor.

Taken together, these observations suggest that neither algorithm exhibits a clear experimental advantage in the $N=15$ regime. Instead, the comparison highlights a trade-off between information concentration and information distribution. Shor’s algorithm benefits from highly structured outputs when noise levels are sufficiently low, while Regev’s algorithm spreads information across multiple correlated outcomes, potentially offering different scaling behavior at larger problem sizes but at the cost of increased sensitivity to noise in higher dimensional sampling.

More broadly, these results emphasize the importance of experimental benchmarking across algorithmic paradigms. Even when two quantum algorithms are asymptotically related in the fault tolerant setting, their behavior on NISQ hardware can differ substantially due to the way quantum information is encoded, propagated, and measured. Understanding these differences is essential for guiding the design and selection of quantum algorithms in the near term.

\begin{table}[t]
\centering
\caption{Circuit depth comparison for Shor’s and Regev’s algorithms for $N=15$ across different execution platforms.}
\label{tab:resources}
\begin{tabular}{lcc}
\hline
Algorithm & Platform & Circuit depth \\
\hline
Shor  & Logical circuit (uncompiled) & $\sim 180$ \\
Shor  & IBM Quantum (ibm\_torino)     & 210 \\
Shor  & QMIO                          & 272 \\
\hline
Regev & Logical circuit (uncompiled) & $\sim 130$ \\
Regev & IBM Quantum (ibm\_torino)     & 147 \\
Regev & QMIO                          & 238 \\
\hline
\end{tabular}
\end{table}

\subsection{Limitations and future work}\label{subsec:limitations}

This study is intentionally limited to the pedagogical benchmark $N=15$,
which enables controlled execution on current hardware but does not
reflect the asymptotic regime where either algorithm becomes
cryptographically relevant. Several limitations follow.

\paragraph{Scaling in $N$.}
Larger moduli require substantially deeper modular arithmetic for Shor
and larger registers or higher dimensional sampling for Regev. Both
effects may dominate the noise behavior observed here, and the relative
trade-off between information concentration and information distribution
may change with problem size. In particular, the distribution based
metrics used in this work may exhibit different scaling properties when
the number of qubits and the circuit depth increase.

\paragraph{Backend dependence and temporal variability.}
The conclusions are conditioned on the specific devices, calibration
states, and transpilation heuristics used. Repeating the benchmark across
additional backends and across multiple calibration snapshots would
strengthen statistical confidence and help separate algorithmic effects
from device specific artifacts. In the NISQ regime, small variations in
hardware noise can significantly modify the observed measurement
distributions.

\paragraph{Sensitivity of classical post-processing.}
Regev’s reconstruction is sensitive to parameter choices (e.g., $R$, the number
of retained outcomes, and lattice reduction settings), while Shor’s
success depends on continued fraction recovery under peak broadening and
phase noise. A systematic sensitivity analysis and a principled selection
of thresholds are left for future work.

Future work will extend the benchmark to larger values of $N$ in
simulation with realistic noise models, increase the sample budget and
dimensionality for Regev’s approach, and explore compilation and error
mitigation strategies (layout, routing, dynamical decoupling, measurement
mitigation) that explicitly target the concentration versus distribution
trade-off highlighted in this paper.
\section*{Use of Generative-AI tools declaration}\label{AI}

The authors used generative-AI tools for language editing and/or formatting assistance. All scientific content,
derivations, and conclusions were produced and verified by the authors, who take full responsibility for the
manuscript.

\section*{Acknowledgments}

This work has been partially supported by the research and development project \emph{``Quantum Shield: Post-quantum Security and Quantum Machine Learning for Cybersecurity in Smart Manufacturing Environments in Industry 4.0''} (Reference CCTT5/23/SA/0002), funded by the Instituto de Competitividad Empresarial of the Junta de Castilla y León (ICECyL) and the European Regional Development Fund (ERDF).

The authors also gratefully acknowledge the computer resources provided by CESGA-QMIO and the technical support received under project IM-2025-2-0044.

\bibliographystyle{unsrt}
\bibliography{Bibliography}

\end{document}